\documentclass{revtex4}
\usepackage{psfig}
\usepackage{subfigure}
\usepackage{graphicx}
\usepackage{doublespace}
\usepackage{amsmath}

\begin{document}

\renewcommand{\baselinestretch}{2}


\doublespace

\title{Network synchronization of groups}
\author{Francesco Sorrentino${}^{\ddagger*}$, Edward Ott${}^{*}$ }
\affiliation{${}^\ddagger$ University of Naples Federico II, Naples 80125, Italy \\ ${}^*$ Institute for Research in Electronics and Applied Physics, Department of Physics, and Department of Electrical and Computer Engineering, University of Maryland, College Park, Maryland 20742}

\begin{abstract}
In this paper we study synchronized motions in complex networks in which there are distinct groups of nodes where the dynamical systems on each node within a group are the same but are different for nodes in different groups. Both continuous time and discrete time systems are considered. We initially focus on the case where two groups are present and the network has bipartite topology (i.e., links exist between nodes in different groups but not between nodes in the same group). 
We also show that group synchronous motions are compatible with more general network topologies, where there are also connections within the groups.
\end{abstract}
\maketitle

\section{Introduction}

Because of its relevance in a wide variety of physical, biological, social and engineering contexts, synchronization of complex networks of
coupled dynamical systems has recently received increasing attention.
In this paper, we analyze synchronized (possibly chaotic) motions in complex networks of 
coupled \emph{groups} of dynamical systems. Here by a group we mean a collection of systems that have the same dynamics, with any given group consisting of systems with dynamics that is different from the dynamics of systems in the other groups. Specifically, we will show that under certain circumstances, multiple group-synchronous evolutions may exist in such networks. In this type of synchronous motion, the evolution of the states of systems within a particular group are the same, while the states of members of different groups although coherently related, are in general different (indeed the state vectors of systems in different groups may have different dimensionality).

The problem of collective behavior in a network connecting members of different groups is of broad interest.
As a first example, we note that 
many efforts have been devoted to the study of teams (groups) of interacting robots performing synchronous coordinated tasks \cite{Jo:Br:Di06,RobotSoccer}. Other studies have regarded the coordination and control of several squadrons (groups) of unmanned autonomous vehicles to accomplish interdependent tasks, such as cooperative searches and attacks  \cite{Ch:Pa:Ra,Fi:Pa:Sp,Passino}.
In the social networks literature, distinct collective behaviors of individuals are often related to their sex, social status and/or race. 
Some studies have clearly pointed out how men's and women's social behaviors differ,
even in situations where they are found to interact tightly, as in virtual communities or internet chats. 
In the brain functional assemblies of neurons have been observed to display distinct interdependent synchronous oscillations \cite{Sc:Gr05}.
Collective dynamics of groups displaying multi-synchronous behaviors have also been uncovered in ecological systems \cite{Blasius:Stone,Bl:Hu:St,Mo:Ku:Bl}, where competition could have favored the evolution of different synchronous behaviors of different species. 
For example, some corals are known to spawn synchronously during a particular season of the year 
\cite{Ryn:Loy}. At the same time, different coral species typically spawn in different months, possibly
to prevent hybridization of the species and/or as a mechanism to relieve
larvae from interspecific competition \cite{St:Ol:Bl:Hu:Ca}.
Distinct roles of males and females (as in the case of social networks) influence the sexual activity of animals, where
reproductive synchrony has been speculated to benefit survival of progeny by decreasing the
likelihood of the male deserting his partner.

%
%

Phase synchronization between essentially different chaotic systems has been the subject of intensive study since the appearance of the paper \cite{Ro:Pi:Ku}. However, here we will be interested in complete (full) synchronization. Moreover, multiple synchronized motions of identical oscillators have been observed to coexist in complex networks characterized by strong community structure \cite{Oh:Rho,Ar:Di:PV,Pa:Lai}. Here we will show that under certain conditions, 
multiple synchronous behaviors of systems with group properties can occur
\footnote{In a recent paper [Q. -C. Pham and J. J. Slotine, Neural networks 20, 62, 2007], the issue of multiple synchronized motions occurring in networks of groups has been independently addressed and the stability of an invariant subspace corresponding to multisynchronicity has been studied for unweighed networks. In our paper, the stability of the multisynchronous evolution is evaluated by introducing a master stability function, which decouples the effects of the network topology from those of the dynamics on the nodes. Moreover, the more general case of weighed networks is considered.}. 

In this paper we focus on the case where there are two groups. In Sec. II we consider bipartite network topology and continuous time dynamics and present examples of both periodic and chaotic synchronous behavior. In Sec. III, discrete time systems are discussed. In the case of bipartite network topology studied in Secs. II and III, a compact master stability function \cite{Pe:Ca} description for evaluating the stability of group synchronous motions is possible. When there are more than two groups or when there are two groups but a non-bipartite network structure, the stability analysis is generally more difficult. In Sec. IV, we remove the constraint of bipartite network topology, and we show that the stability of the multi-synchronous evolutions is indeed possible under these more general conditions and that it can be enhanced when connections are allowed between systems belonging to the same group.

\section{Continuous time bipartite systems}

In this section we focus on continuous time systems and consider a bipartite network connecting two groups. We find the conditions that allow  a synchronization manifold and study its stability by means of a master stability function approach.

\subsection{Formulation}

The individual
equation of an isolated (uncoupled) node is denoted by $\dot x_i=F(x_i)$,
$i=1,...,N_x$, for the nodes in the first group $\mathcal{S}_x$ and by
$\dot y_j=G(y_j)$ $j=1,...,N_y$ for the nodes in the second group
$\mathcal{S}_y$, where $x_i$ ($y_j$) is an $n_x$-dimensional ($n_y$-dimensional) state vector and $F: \mathbf{R}^{n_x} \rightarrow  \mathbf{R}^{n_x}$ and $G: \mathbf{R}^{n_y} \rightarrow  \mathbf{R}^{n_y}$.
The dynamical equations of the network systems are as follows:
\begin{equation}
\begin{split}
\dot x_i= F(x_i) + \sum_{j=1}^{N_y} A_{ij} H(y_j), \quad i=1,...,N_x, \label{sys} \\
\dot y_j= G(y_j) + \sum_{i=1}^{N_x} B_{ji} L(x_i), \quad j=1,...,N_y,
\end{split}
\end{equation}
where $A$ is an $N_x \times N_y$ coupling matrix, whose entries $\{ A_{ij} \}$ represent the intensity of the direct interaction
from system  $j$ in $\mathcal{S}_y$ to $i$ in $\mathcal{S}_x$. Analogously the entries $\{ B_{ji} \}$ of the $N_y \times N_x$ matrix $B$ represent the interaction
from system  $i$ in $\mathcal{S}_x$ to $j$ in $\mathcal{S}_y$. The interaction function $H$ ($L$) is a mapping from $\mathbf{R}^{n_y}$ to $\mathbf{R}^{n_x}$ (from $\mathbf{R}^{n_x}$ to  $\mathbf{R}^{n_y}$). 

We now consider the possibility of the existence of multi-synchronous solutions, where by this we mean that $x_1(t)=x_2(t)=...=x_{N_x}(t)= {x}_{s}(t)$
and $y_1(t)=y_2(t)=...=y_{N_y}(t)= {y}_{s}(t)$. Substituting such an assumed solution in (\ref{sys}), we see that in order for a multi-synchronous state to exist the sum $\sum_j A_{ij}$ must be independent of $i$ and the sum $\sum_i B_{ji}$ must be independent of $j$. If we denote the first sum by $a$ and the second sum by $b$, then by the replacements $a H \rightarrow H $ and $A/a \rightarrow A$ ($b L \rightarrow L$ and $B/b \rightarrow B$) we see that, without loss of generality, it suffices to set $a=b=1$,

\begin{subequations}
\begin{align}
  \sum_{j=1}^{N_y} A_{ij}=1 \quad \forall i \in \mathcal{S}_x,  \label{lim1} \\
  \sum_{i=1}^{N_x} B_{ji}=1 \quad \forall j \in \mathcal{S}_y. \label{lim2}
\end{align}
\end{subequations}

Thus the equations of motion for the synchronized dynamics are

\begin{equation} \begin{split}
\dot {x_s}= F(x_s) + H(y_s), \label{star} \\
\dot {y_s}= G(y_s) + L(x_s).
\end{split} \end{equation}

\subsection{Synchronization Stability}

In what follows we seek to characterize the stability of the above defined synchronous state. Linearization of the system (\ref{sys}) around the synchronous evolutions $x_s(t)$ and $y_s(t)$ yields:
\begin{equation} \begin{split}
\dot {\delta x_i}= DF(x_s) \delta x_i + \sum_{j=1}^{N_y} A_{ij} DH(y_s) \delta y_j, \quad i=1,...,N_x, \label{linsys} \\
\dot {\delta y_j}= DG(y_s) \delta y_j + \sum_{i=1}^{N_x} B_{ji} DL(x_s) \delta x_i, \quad j=1,...,N_y .
\end{split} \end{equation}

The Lyapunov exponents of the dynamics of a synchronous state ($x_s(t), y_s(t)$) are those associated with the follow system:
\begin{equation}
\begin{split}
\dot {\delta x_s}= DF(x_s) \delta x_s + DH(y_s) \delta y_s,  \\
\dot {\delta y_s}= DG(y_s) \delta y_s + DL(x_s) \delta x_s, \label{synsys}
\end{split}
\end{equation}
obtained by linearization of Eqs. (\ref{star}).
Note that the synchronous evolutions $x_s$ and $y_s$ might, e.g., be stationary, periodic, or chaotic.

 We now assume that the $(N_x + N_y)$ independent solutions of (\ref{linsys}) can be expressed in the form $\delta x_i= c_{x_i} \delta \bar{x}$, $i=1,...,N_x$ and $\delta y_j = c_{y_j} \delta \bar{y}$, $j=1,...,N_y$, where $\{ c_{x_i} \}$ and $\{ c_{y_j} \}$ are appropriate time-independent scalars. 
 This assumed form will encompass all possible linear solutions of (4) if the space of vectors given by the possible values of $c_{x_i}, c_{y_j} (i=1,...,N_x; j=1,...,N_y)$ has dimension $N_x+N_y$. As we shall see, this is the case (cf. Eq. (9) to follow).
 With the assumption, $\delta x_i= c_{x_i} \delta \bar{x}$, $i=1,...,N_x$ and $\delta y_j = c_{y_j} \delta \bar{y}$, $j=1,...,N_y$,  Eqs. (\ref{linsys}) become
\begin{subequations} \begin{align}
c_{x_i}  {\delta \dot{ \bar{x}}}= c_{x_i} DF(x_s) \delta \bar{x} + (\sum_{j=1}^{N_y} A_{ij} c_{y_j}) DH(y_s) \delta \bar{y}, \quad i=1,...,N_x \label{linsys2a} \\
c_{y_j}  {\delta \dot{ \bar{y}}}= c_{y_j} DG(y_s) \delta \bar{y} + (\sum_{i=1}^{N_x} B_{ji} c_{x_i}) DL(x_s) \delta \bar{x}, \quad j=1,...,N_y . \label{linsys2b}
\end{align} \end{subequations}

Thus in order that (\ref{linsys2a}), (respectively (\ref{linsys2b})), is satisfied for all $i$, (respectively $j$), we require that 
${c_{x_i}^{-1}}{\sum_j A_{ij} c_{y_j}}= \nu$, where $\nu$ is independent of $i$
and ${c_{y_j}^{-1}}{\sum_i B_{ji} c_{x_i}}= \eta$, where $\eta$ is independent of $j$. After defining the vectors $c_x=(c_{x_1}, c_{x_2},..., c_{x_{N_x}})^T$ and $c_y=(c_{y_1}, c_{y_2},..., c_{y_{N_y}})^T$, these conditions may be rewritten as: $A c_y= \nu c_x$ and $B c_x = \eta c_y$; that is,

\begin{equation}
\left(
  \begin{array}{cc}
    0 & A \\
    B & 0 \\
  \end{array}
\right) \left( \begin{array}{c}
          c_x \\
          c_y
        \end{array} \right) = \left( \begin{array}{c}
                      \nu c_x \\
                      \eta c_y
                    \end{array} \right). \label{OABO}
\end{equation}

Using this in (5) we obtain
\begin{equation} \begin{split}
{\delta \dot{ \bar{x}}}= DF(x_s) \delta \bar{x} + \nu DH(y_s) \delta \bar{y},  \label{mueta} \\
{\delta \dot{ \bar{y}}}= DG(y_s) \delta \bar{y} + \eta DL(x_s) \delta \bar{x}.
\end{split} \end{equation}

One particular solution of (\ref{OABO}) is obtained when $\nu=\eta=\lambda$, i.e.,

 \begin{equation}
Q \left( \begin{array}{c}
          c_x^0 \\
          c_y^0
        \end{array} \right) = \lambda \left( \begin{array}{c}
                       c_x^0 \\
                       c_y^0
                    \end{array} \right), \qquad \qquad Q=\left(
  \begin{array}{cc}
    0 & A \\
    B & 0 \\
  \end{array}
\right),  \label{sette}
\end{equation}
where $\lambda$ belongs to the set $\Lambda=\{ \lambda_i \}$, $i=1,...,N_x+N_y$ of the (possibly complex) eigenvalues of the matrix $Q$.

%

Rewriting (\ref{sette}) as
 \begin{equation}
\left(
  \begin{array}{cc}
    0 & A \\
    B & 0 \\
  \end{array}
\right) \left( \begin{array}{c}
          c_x^0 \\
          z c_y^0
        \end{array} \right) = \left( \begin{array}{c}
                       (\lambda z) c_x^0 \\
                       (\lambda/z) z c_y^0
                    \end{array} \right), \label{nove}
\end{equation}
shows that solution of (\ref{sette}) yields all the possible solutions of (\ref{OABO}) by setting $\nu=\lambda z$, $\eta= \lambda/z$, $c_x=c_x^{0}$, $c_y=z c_y^{0}$, where $z$ is a free parameter. Also since $A$ and $B$ are real, the spectrum of $Q$ is symmetric about the $Re(\lambda)$  axis. Furthermore, we note that, if $\lambda$ is an eigenvalue of $Q$, then, by letting $z=-1$, we see that $-\lambda$ is also an eigenvalue. Thus the spectrum of $Q$ is symmetric about both the $Re(\lambda)$ axis as well as the $Im(\lambda)$ axis.

Moreover, the stability of the synchronous evolutions does not depend on the particular $z$. In fact, if in Eqs. (\ref{mueta}) we let $\nu= \lambda z$, $\eta= \lambda/z$, $\delta \tilde{y} = z \delta\bar{y}$, we see that $\delta \bar{x}, \delta \tilde{y}$ satisfy Eqs. (\ref{mueta}) with $\nu=\eta=\lambda$. Thus it suffices to consider (\ref{sette}), and we rewrite Eqs. (\ref{mueta}) as

\begin{equation}
\begin{split}
\delta \dot{\bar{x}}= DF(x_s) \delta \bar{x} + \lambda DH(y_s) \delta \bar{y}, \\
\delta \dot{\bar{y}}= DG(y_s) \delta \bar{y} + \lambda DL(x_s) \delta \bar{x}, \label{BLOCK}
\end{split}
\end{equation}
where $\lambda=\lambda_1,\lambda_2,...,\lambda_N$.
  We define a master stability function \cite{Pe:Ca} for this problem, denoted $M(\lambda_i)$, where $M$ associates to $\lambda$, the maximum Lyapunov exponent of the system (\ref{BLOCK}). Note that the function $M(\lambda)$ can be determined without knowledge of the matrix $Q$. Thus the synchronization stability problem is decomposed in two parts, (i) a part dependent only on the couplings $H$ and $L$ and on the individual system dynamics $F$ and $G$, but not on the network topology (i.e. not on the matrix $Q$), and (ii) a part dependent solely on the network topology (determination of the spectrum of $Q$).

Another important consequence of the invariance of (\ref{mueta}) under the transformation $\nu \rightarrow \lambda z$, $\eta \rightarrow \lambda/z$ is that the synchronous state stability for an eigenvalue $\lambda$ is the same as for $-\lambda$ ($z \rightarrow -z$). Thus only those eigenvalues with, e.g., $Re(\lambda) \geq 0$ need to be tested. 


\subsection{Spectrum of $Q$}

The matrix $Q$ has a pair of real eigenvalues $1$ and $-1$. This follows because the sums of the components for all rows of $A$ and $B$ are one. The eigenvalue $+1$ corresponds to an eigenvector all of whose components have the same value; while the eigenvector $-1$ corresponds to an eigenvector whose first $N_x$ components have the same value and whose remaining $N_y$ components all have the negative of this value. Hence the eigenvalues $+1$ and $-1$ are associated with the directions parallel to the synchronization manifold; thus they may result in positive Lyapunov exponents corresponding to chaotic dynamics taking place in the synchronization manifold $(x_1=x_2=...=x_{N_x},y_1=y_2=...=y_{N_y})$. In order to check the stability of the synchronous evolutions, one should evaluate the master stability function $M(\lambda_i)$  for the remaining $N_x+N_y-2$ eigenvalues of $Q$, representing the stability of the motions transverse to the synchronization manifold. The synchronized state is stable if, $M(\lambda_i)<0$ for all $\lambda_i$ in the set $\Lambda'=\{\Lambda-\{-1,+1\}\}$.

From the fact that the sum of the elements in every row of $Q$ is one, with all zero elements on the main diagonal, the Gershgorin circle theorem implies that the spectrum of $Q$ lies in the disc of unit radius in the complex plane, having its center at ($0,0$).
Note also that the matrix $Q$ has at least $|N_x-N_y|$ zero eigenvalues, so that zero eigenvalues must always occur unless $N_x=N_y$. In particular, if $N_x \neq N_y$, a necessary condition for the stability of the synchronized coupled systems is that the Lyapunov exponents resulting from $\delta \dot{\bar{x}}= DF(x_s) \delta \bar{x}$ and  $\delta \dot{\bar{y}}= DG(y_s) \delta \bar{y}$ are all negative. (Note that these exponents depend on $H$ and $L$ because the synchronous time evolutions $x_s(t)$ and $y_s(t)$ depend on $H$ and $L$.)
In order to see that $Q$ has at least $|N_x-N_y|$ zero eigenvalues, assume that $N_x > N_y$. Then the $N_x$ rows of $A$ (each of which has $N_y <  N_x$ components) can span a space of at most dimension $N_y$. Hence the $N_x+N_y=N$ rows of $Q$ can span a space of most dimension $2 N_y$, and there are at least $(N_x+N_y)-2 N_y=N_x-N_y$ independent homogeneous linear relationships between the rows of $Q$, implying that there are at least $N_x-N_y$ zero eigenvalues.

Moreover, the spectrum of $Q$ can be obtained through the computation of the eigenvalues of the lower dimensional of the two matrices, $AB$ and $BA$.  In fact, by noticing that $Q^2$ is a block diagonal matrix of the form
 \begin{equation}
Q^2= \left(
  \begin{array}{cc}
    AB & 0 \\
    0  & BA \\
  \end{array}
\right), \label{ABBA}
\end{equation}
we have that, if $\lambda$ is in the spectrum of $Q$, then $\lambda^2$ must be one of the $N_x$ eigenvalues of $AB$ and/or one of the $N_y$ eigenvalues of $BA$. Say $N_{min}= \min{(N_x,N_y)}$, define the $N_{min} \times N_{min}$ matrix,
\begin{align}
D= \left\{ \begin{array} {ccc} {AB,} \quad \mbox{if} \quad {N_x \leq N_y,} \\ {BA,}  \quad \mbox{if} \quad    {N_y < N_x,} \end{array} \right. 
\end{align}
and denote the spectrum of $D$ by
$\tilde \Lambda=\{ \tilde {\lambda}_1,..., \tilde \lambda_{{N}_{min}} \}$.
Then, since the eigenvalues of $Q^2$ are the square of the eigenvalues of $Q$, we have that the spectrum of $Q$ is
\begin{equation}
\Lambda= [0,0,...,0] \bigcup [\pm \sqrt{\tilde{\lambda}_1}, \pm \sqrt{\tilde{\lambda}_2},...,\pm \sqrt{\tilde \lambda_{{N}_{min}}}], \label{quattordici}
\end{equation}
where $[0,0,...,0]$ denotes $|N_x-N_y|$ zeros. (Note that by Eqs. (2) one of the eigenvalues of $D$ is $+1$, corresponding to an eigenvector $[1,1,...,1]^T$.)

\begin{figure}
\centering
\begin{minipage}[c]{1\textwidth}
\centering
\includegraphics[width=1\textwidth]{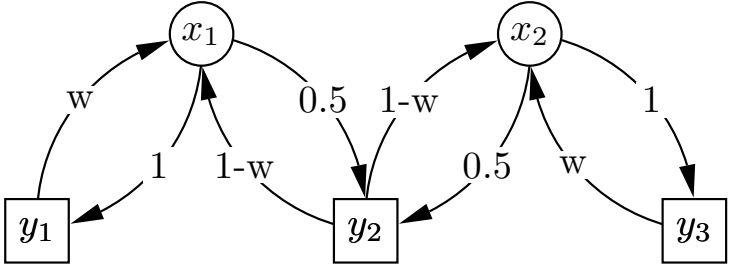}
\caption{A simple 5-nodes network.} \label{5Nb}
\end{minipage}%
\end{figure}

As an example, we consider the network in Fig. \ref{5Nb}. For this network $N_x=2$, $N_y=3$, and
\begin{equation}
A= \left(
     \begin{array}{ccc}
       w & 1-w & 0\\
       0 & 1-w & w \\
     \end{array}
   \right), \quad B= \left(
                       \begin{array}{cc}
                         1 & 0 \\
                         1 \over 2 & 1 \over 2 \\
                         0 & 1 \\
                       \end{array}
                     \right).
\end{equation}
Note that the row sums of $A$ and $B$ are one as required by Eqs. (\ref{lim1},\ref{lim2}). Since $N_x=2 <N_y=3$, $D=AB$, so that
\begin{equation}
D= \frac{1}{2}\left(
     \begin{array}{cc}
       1+w & 1-w \\
       1-w & 1+w \\
     \end{array}
   \right).
\end{equation}

The eigenvalues of this $2 \times 2$ matrix are $1$ and $w$. Thus since $|N_x-N_y|=1$, Eq. (\ref{quattordici}) yields the real spectrum,
\begin{equation}
\Lambda=[-1, -\sqrt{w},0,\sqrt{w},1]. \label{diciassette}
\end{equation}

We now consider the spectrum of $Q$ for large networks of two types: (i) $Q$ is random, and (ii) $Q$ is constrained to have a real spectrum but is otherwise random. To construct the matrix $Q$ in these two cases we start with a matrix $Q'$ of the form
 \begin{equation}
Q'= \left(
  \begin{array}{cc}
    0 & A' \\
    B' & 0 \\
  \end{array}
\right), \label{X}
\end{equation}
and then take $Q$ to be
\begin{equation} Q=K^{-1} Q' \label{Y} \end{equation}
where $K$ is a diagonal matrix with $K_{ii}= \sum_j Q'_{ij}$. Equation (\ref{Y}) insures that the row sums of $Q$ are one as required by Eqs. (\ref{lim1},\ref{lim2}). For case (i) we choose the elements of $A'$ ($B'$) randomly to be one with probabilities $p_{xy}$ ($p_{yx}$), and zero otherwise. Note that, in case (i) there is no correlation between $A_{ij}$ and $B_{ji}$. For case (ii) we choose $A'$ randomly with $A'_{ij}=1$ with probability $p$ and $A'_{ij}=0$ otherwise, and we then set $B'=(A')^T$. Thus in this case, $Q'$ is symmetric and the first $N_x$ elements $K_{ii}$ of $K$ are the row sums of $A'$, while the next $N_y$ components are the column sums of $A'$. For case (ii) the spectrum is real, since by multiplying by $K^{1/2}$, the eigenvalue equation $Q c = \lambda c$ can be rewritten as $(K^{-1/2} Q' K^{-1/2}) c'= \lambda c'$, where $c'= K^{1/2} c$. Because $K^{-1/2} Q' K^{-1/2}$ is symmetric, if $Q'$ is, we see that the eigenvalues of $Q$ are real in case (ii). Next we use numerical experiments to investigate the general properties of the spectrum of large random matrices $Q$ in the above two cases.

\begin{figure}[t]
\centerline{
\psfig{figure=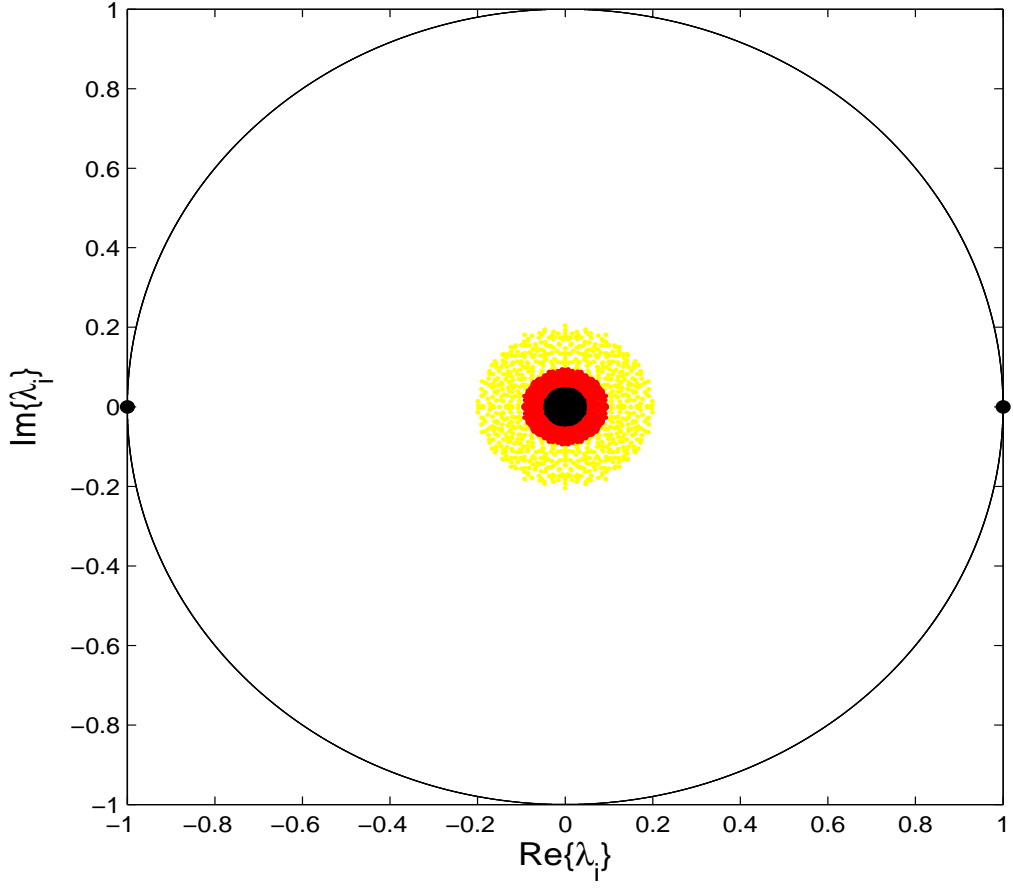,width=15cm,height=13cm}}
\caption{(Color online) \small Random networks with $N_x=N_y=5 \times 10^2$ nodes. The spectrum of $Q$ is shown for three sets of values of $p_{xy}$ and $p_{yx}$.
Yellow (light gray) is used for $p_{xy}=p_{yx}=0.05$, red (dark gray) for $p_{xy}=0.5$ and $p_{yx}=0.05$, black for $p_{xy}=p_{yx}=0.5$. The continuous line is used to represents the circle of unit radius centered at (0,0). The eigenvalues $\lambda = \pm 1$ associated with perturbations in the synchronization manifold are denoted by solid black dots.  \label{BERS}}
\end{figure}

\begin{figure}[h]
\centerline{
\psfig{figure=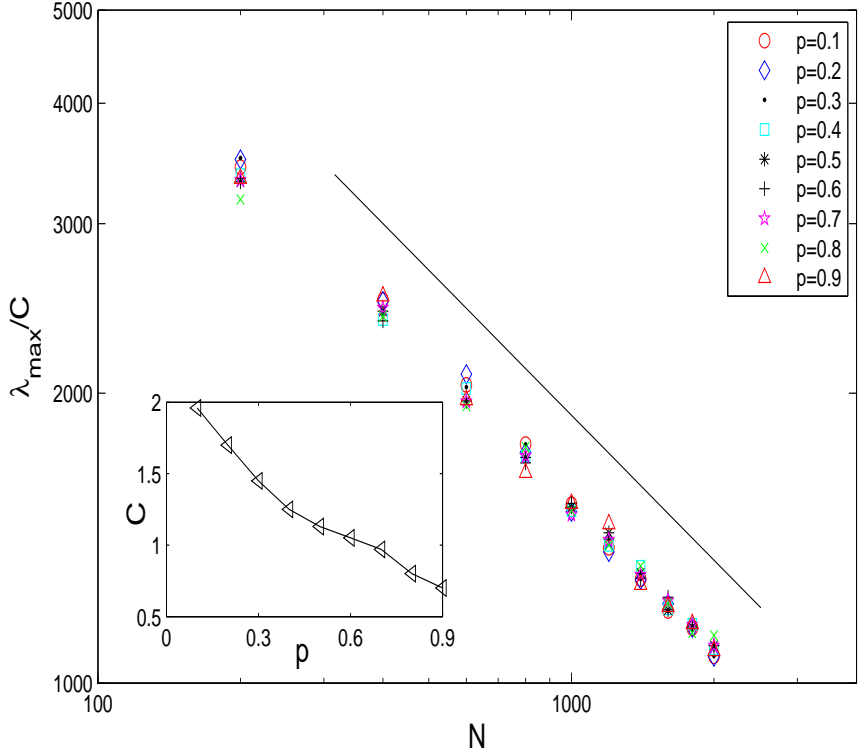,width=13cm,height=11cm}}
\caption{(Color online) \small log-log plot of $\lambda_{max}/C$ versus $N$, for different values of $p_{xy}$ ranging between $0.1$ and $0.9$ in steps of 0.1.  The straight line has slope $-1/2$. The inset shows $C$ versus $p$. \label{SC3}}
\end{figure}

First we consider case (i). We take $N_x=N_y=500$ and find the spectrum of $Q$ for randomly generated matrices with several values of $p_{xy}$ and $p_{yx}$. Results are shown in Fig. \ref{BERS}. We see that there are two eigenvalues at $\lambda= \pm 1$ and that all the other eigenvalues lie within a circle whose radius decreases as the average node degree increases (i.e., as $p_{xy}$ and $p_{yx}$ increase). 
We evaluated the scaling of  $\lambda_{max}=\max_i | {\lambda_i} |$ for $\lambda_i \in \Lambda'$ with the network size $N$ in the simple case where $N_x=N_y=N/2$ and $p_{xy}=p_{yx}\equiv p$. We hypothesize a scaling of the form $\lambda_{max}(N,p_{xy})= C N^{\epsilon}$, and
perform numerical simulations with $p$ ranging between $0$ and $1$, and $N$ ranging between $200$ and $2000$. Our numerics show that $\epsilon \simeq 1/2$ (we note, however, that for $p_{xy} \rightarrow 1$, $\lambda_{max}=0$, independent of $N$).
Thus, by assuming a scaling of the form $\lambda_{max}(N,p)= C N^{-1/2}$ we obtained different values for $C$, as function of the probability $p$.
In Fig. \ref{SC3} the values of the logarithm of $\lambda_{max}/C$  are shown to collapse to a straight line of slope $-1/2$ for different values of $p$, when plotted versus the logarithm of the network dimension $N$. The inset of Fig.\ref{SC3} shows $C$ versus $p$ for $p$ ranging between $0$ and $0.9$ in steps of $0.1$. The scaling $\lambda_{max} \sim N^{-1/2}$ implies that, 
with increasing $N$, the spectrum $\Lambda'$ shrinks toward the point $(0,0)$. 
Moreover, $\Lambda'$ also shrinks toward $(0,0)$ as $p_{xy}$ and $p_{yx}$ approach one, independently of the network dimension $N$.  Thus both in the case of a very large network (i.e., $N$ large) or complete network (i.e., $p_{xy}, p_{yx} \rightarrow 1$), the whole spectrum of the eigenvalues in $\Lambda'$ collapses onto the real eigenvalue $0$.

%

\begin{figure}[t]
\centerline{
\psfig{figure=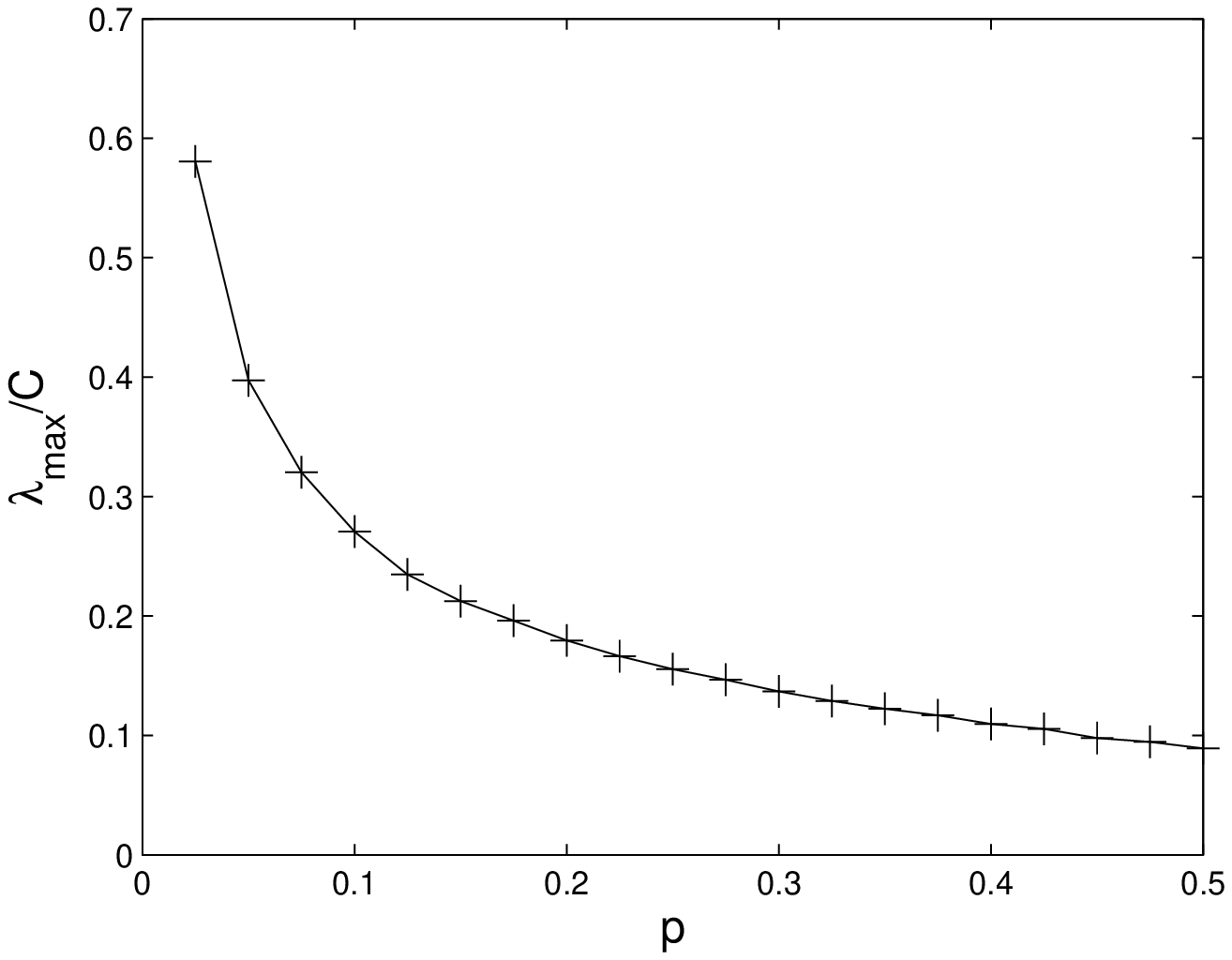,width=9cm}\psfig{figure=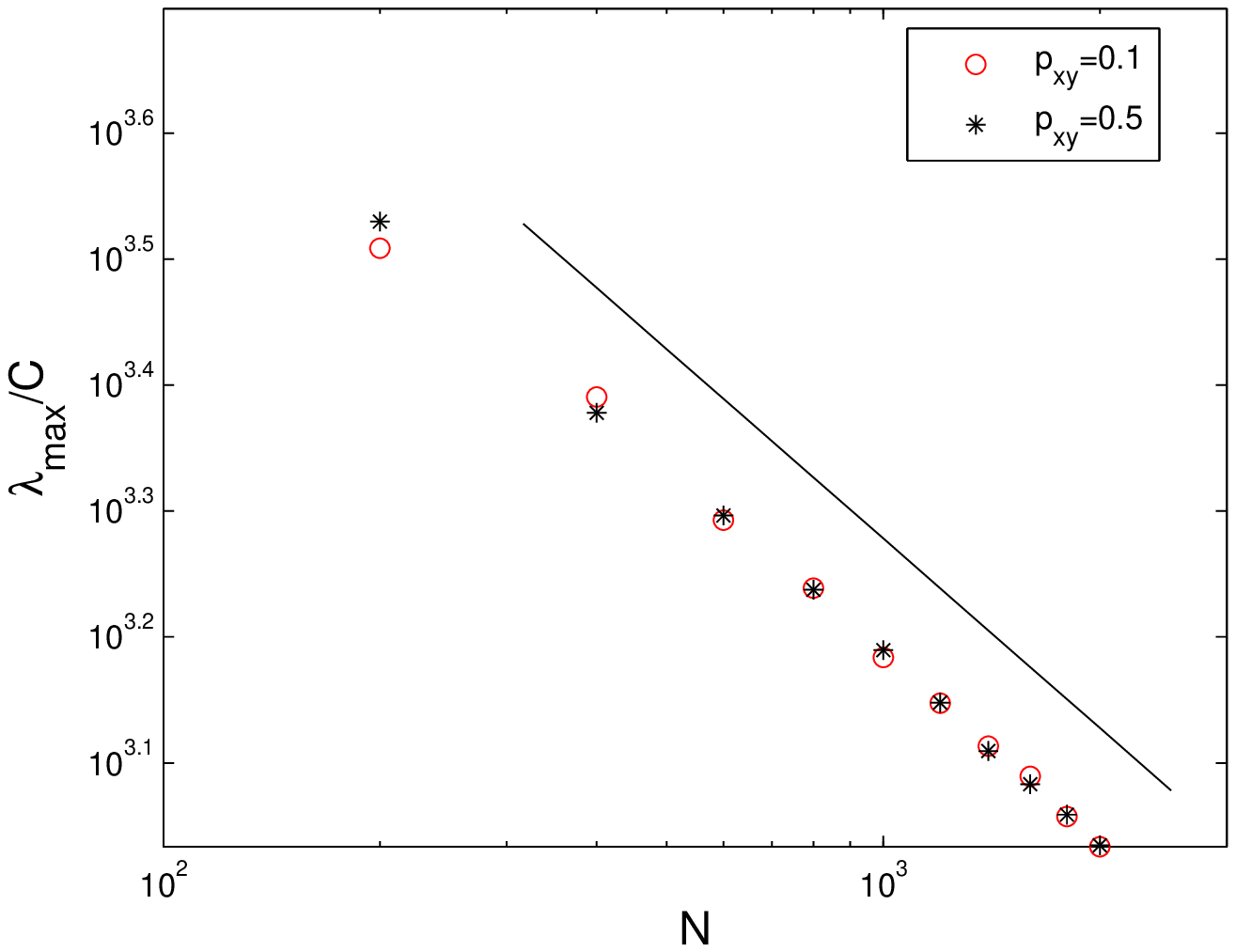,width=9.1cm}}
\caption{ \small The behavior of $\lambda_{max}$ as function of $p$ and of $N$ for large random networks with real spectra ($Q=K^{-1} Q'$ and $Q'$ symmetric). (a) $\lambda_{max}$ versus $p$, for $N_x=N_y=5 \times 10^2$ and $N_y=5 \times 10^2$ nodes. (b) Log-log plot of $\lambda_{max}/C$ versus $N$ for $p=0.1$ and $p=0.5$ showing that (as for case (i)) $\lambda_{max}$ scales as $N^{-1/2}$ for large $N$ (the solid line has slope $-1/2$). \label{BERS2}}
\end{figure}%

%

We now consider case (ii) where the spectrum of $Q$ is real. Analogous to the results in Figs. \ref{BERS} and \ref{SC3}, we find that for large $N$ the eigenvalues of $Q$ in $\Lambda'$ are distributed along the real line lying in a symmetric range, $-\lambda_{max} \leq \lambda \leq \lambda_{max}$, where $\lambda_{max}$ decreases toward zero with increasing $p$ (see Fig. \ref{BERS2}(a)) as well as increasing $N$ (see Fig. \ref{BERS2}(b)), with $\lambda_{max} \sim N^{-1/2}$ for large $N$.

The result that $\lambda_{max}$ decreases with $N$ and $p$ (or $p_{xy}$ and $p_{yx}$) for both cases (i) and (ii) is quite significant. In particular, if $\lambda_{max} \ll 1$, then, for the purposes of evaluating the master stability function, it becomes a good approximation to set $\lambda=0$. This is a great simplification in that the master stability function now need be evaluated only for a single value of $\lambda$, and its determination  reduces to a computation on two uncoupled systems,
\begin{equation}
\begin{split}
\delta \dot{\bar{x}}= DF(x_s) \delta \bar{x},  \\
\delta \dot{\bar{y}}= DG(y_s) \delta \bar{y}, \label{simpl}
\end{split}
\end{equation}
where we again emphasize that, although the coupling functions $H$ and $L$ do not appear explicitly in (\ref{simpl}), $M$ still depends on $H$ and $L$ because the synchronous time evolutions, $x_s(t)$ and $y_s(t)$, depend on $H$ and $L$ (see Eq. (\ref{star}) ).

\subsection{Examples}

\textit{Example 1: Synchronized Periodic Motion.}

We consider the following coupled network dynamical equations, that are in the form (\ref{sys}),
\begin{align}
\dot {x}_{i(1)}= x_{i(2)} -x_{i(1)} ({x_{i(1)}^2}+{x_{i(2)}^2}-1) +& \sigma_x \sum_{j=1}^{N_y} A_{ij} y_{j(1)}, \nonumber\\
\dot {x}_{i(2)}= -x_{i(1)} -x_{i(2)} ({x_{i(1)}^2}+{x_{i(2)}^2}-1),& \qquad \qquad i=1,...,N_x. \label{esemx}
\end{align}
\begin{align}
\dot {y}_{j(1)}= y_{j(2)} + \sigma_y \sum_{i=1}^{N_x} B_{ji} x_{i(1)},& \nonumber\\
\dot {y}_{j(2)}= -y_{j(1)} -0.2 y_{j(2)} ({y_{j(1)}^2}-1),& \qquad \qquad j=1,...,N_y.\label{esemy}
\end{align}
In the absence of coupling $\sigma_x=\sigma_y=0$, Eqs. (\ref{esemx}) and (\ref{esemy}) both individually have global attractors on which the motion is periodic (i.e., they are limit cycle attractors). In particular, with $\sigma_y=0$,  Eq. (\ref{esemy})  is the Van der Pol equation. 
%

In order to measure the extent to which synchronization is achieved,  we have monitored the asymptotic time average of the following two quantities: $E_x=\frac{1}{N_{x}^2} \sum_{i=1}^{N_x} \sum_{j=1}^{N_x} (|{x}_{i(1)}-{x}_{j(1)}|+|{x}_{i(2)}-{x}_{j(2)}|)$ and $E_y=\frac{1}{N_{y}^2} \sum_{i=1}^{N_y} \sum_{j=1}^{N_y} (|{y}_{i(1)}-{y}_{j(1)}|+|{y}_{i(2)}-{y}_{j(2)}|)$, as functions of the control parameter $\sigma_x$ with $\sigma_y=0.65$. For each $i$ and $j$ we consider randomly chosen initial conditions in $|x_{i1,2}|<3$ and $|y_{j1,2}|<3$ and evolve the system for a long time (from $t=0$ to $t=300$).


%
%

\begin{figure}[h]
\centerline{
\psfig{figure=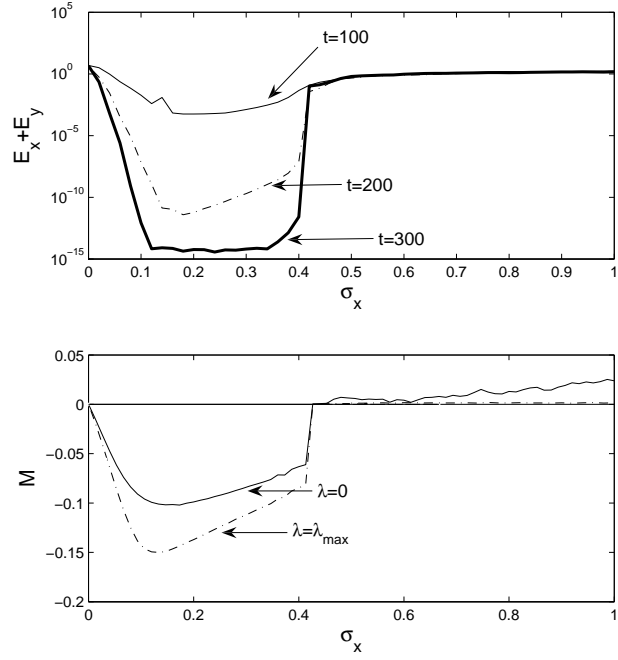,width=9cm}}
\caption{\small The upper panel shows ${E_x+E_y}$ versus $\sigma_x$ for $\sigma_y=0.65$. The continuous thin line represents $E_x+E_y$ at $t= 100$; the dashed line, $E_x+E_y$ at $t=200$; the thick continuous line, $E_x+E_y$ at $t=300$. The network parameters are as follows:  $N_x=200$ and $N_y=300$, and $p=0.5$. The lower panel shows the master stability function evaluated at $\lambda=0$ (continuous line), and at $\lambda=\lambda_{max}=0.13$ (dashed line) versus $\sigma_x$. \label{SPA}}
\end{figure}

The case of a network with a real spectrum, obtained as explained in Sec. II.C for the case (ii), is shown in Fig. \ref{SPA}. For this network we take $N_x=200$, $N_y=300$ and $p=0.5$, for which we find $\lambda_{max}=0.13$. The upper panel of Fig. \ref{SPA}, shows $E_x+E_y$ as functions of $\sigma_x$ for $\sigma_y=0.65$ at different simulation times $t=100,200,300$. We see that for $0 < \sigma_x \leq 0.4$ the error decreases with time to very low values, indicating stable synchronization in this range of $\sigma_x$ (the synchronized motion in this range is observed to be periodic). The lower panel shows the corresponding master stability function evaluated at $\lambda=0$ (continuous line), and at $\lambda=\lambda_{max}=0.13$ (dashed line). We observe that the $\sigma_x$ value for the zero crossing of $M(\lambda)$ is approximately at $0.4$, for both $\lambda=0$ and $\lambda=\lambda_{max}$. For other values of $\lambda$ in the range $0 < \lambda < \lambda_{max}$ the curves are similar and have $\sigma_x$ values at the zero crossings of the master stability function at approximately $0.4$. 
Thus we find that the master stability function (lower panel of Fig. \ref{SPA}) predicts a stable $\sigma_x$ range of $0 \leq  \sigma_x  \leq 0.4$ in excellent agreement with our results from the full nonlinear computation (upper panel of Fig. \ref{SPA}). 

\textit{Example 2: Synchronized Chaotic Motion.}

We now consider the following coupled network dynamical equations,
\begin{align}
\dot {x}_{i}= -r (x_{i} + h(x_{i}) )  +r \sum_{j=1}^{N_y} A_{ij} y_{j(1)}, \qquad \qquad i=1,...,N_x, \label{esemcx}
\end{align}
\begin{align}
\dot {y}_{j(1)}= -y_{j(1)}+y_{j(2)} +  \sum_{i=1}^{N_x} B_{ji} x_{i},& \nonumber\\
\dot {y}_{j(2)}= -q y_{j(1)}, & \qquad \qquad j=1,...,N_y,\label{esemcy}
\end{align}
where $h(x)= m_1 x + \frac{m_0-m_1}{2} (|x+1|-|x-1|)$ and we take $r=4.6$, $q=6.02$, $m_0=-8/7$, $m_1=-5/7$.

When $x_i=x_s \forall i$, $y_{j(1,2)}=y_{s(1,2)} \forall j$, the three equation system formed by (\ref{esemcx}) and (\ref{esemcy}) has three attractors; two are stable fixed points at $(x,y_{(1)},y_{(2)})=(\pm 3/2,0, \mp 3/2)$ and the third is a chaotic attractor \cite{ChenBOOK}. Thus, depending on the initial conditions, the motion in the synchronization manifold can be chaotic. 
We now investigate the stability of the synchronous chaotic motions for large $N$.
 For $N>>1$ all the eigenvalues in $\Lambda'$ tend to $0$, and the synchronous evolution thus is stable if the maximum Lyapunov exponents associated with the following two (uncoupled) systems,
\begin{align}
\delta \dot {x}= K(t) \delta x, \quad \mbox{where} \quad K(t)=  -r- \left\{ \begin{array} {ccc} {r m_0,} \quad \mbox{if} \quad {|x_s|<1}, \\ {r m_1,}  \quad \mbox{if} \quad    {|x_s|>1}, \end{array} \right. \label{esemc2x} 
\end{align}
and
\begin{equation}\begin{split}
\delta \dot {y}_{(1)}= -\delta y_{(1)}+\delta y_{(2)}, \\
\delta \dot {y}_{(2)}= -q \delta y_{(1)}, \label{esemc2y}
\end{split}\end{equation}
are both negative. Note that the $x$-Lyapunov exponent for the system in (\ref{esemc2x}) is the time average of $K(t)$ which is equal to $-r (1+p_{<} m_o + p_{>} m_1)$, where $p_{<}$ ($p_{>}$) is the fraction of time that $|x_s(t)| <1$ $(|x_s(t)| >1)$, where $p_{<} + p_{>} \equiv 1$. Hence we have that the $x$- Lyapunov exponent is negative if $p_{<} < (1+m_1) / (m_1 -m_0)=2/3.$ From numerical solution for the synchronized motion we find that this condition is indeed satisfied.
 On the other hand, the system (\ref{esemc2y}) is equivalent to $ \ddot s = - \dot s - q s$, where $s=\delta y_{(2)}$, which converges toward the origin $(0,0)$ with Lyapunov exponents, both equal to $-1/2$.
Thus the synchronization of the network in (\ref{esemcx}) and (\ref{esemcy}), is ensured for sufficiently large networks.

\begin{figure}[h]
\centerline{\psfig{figure=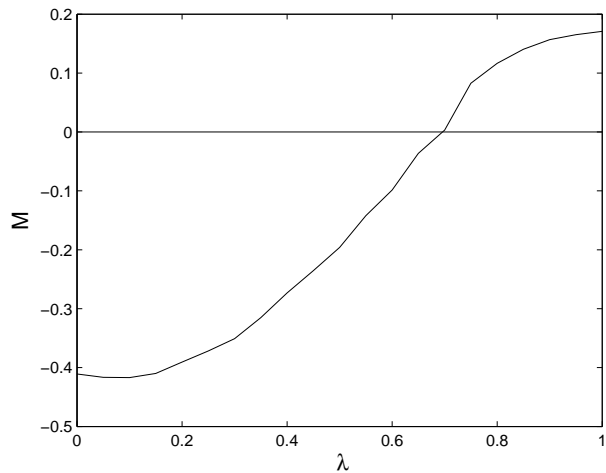,width=9cm}}
\caption{\small The master stability function associated with the system (\ref{esemcll}) as function of the real parameter $\lambda$, varying between $0$ and $1$. \label{MSFChua}}
\end{figure}
We now investigate synchronization stability for the systems (\ref{esemcx}) and (\ref{esemcy}) for a case of a real spectrum for $Q$, but without assuming large $N$. By linearizing the system in (\ref{esemcx}) and (\ref{esemcy}) about the synchronous evolution, we obtain for Eqs. (\ref{BLOCK}):
%
\begin{equation}
\frac{d}{dt} \left( \begin{array} {c} \delta x \\ \delta y_{(1)} \\ \delta y_{(2)} \end{array} \right) = \left( \begin{array} {c c c}  K(t) & \lambda r & 0 \\ \lambda & -1 & -1 \\ 0  & -6.02 & 0      \end{array} \right) \left( \begin{array} {c} \delta x \\ \delta y_{(1)} \\ \delta y_{(2)} \end{array} \right). \label{esemcll}
\end{equation}
In Fig. \ref{MSFChua}, we evaluate the master stability function associated with the system in (\ref{esemcll}) as function of $\lambda$.
The figure shows that, if all $\lambda_i$ in $\Lambda'$ lie in the range $(0,0.7)$, synchronization will be stable. Moreover, since the master stability function (in Fig. \ref{MSFChua}) becomes positive as $\lambda$ increases, the stability of the synchronous evolution depends only on $\lambda_{max}$; i.e., if $M(\lambda_{max})$  is negative, the synchronous evolution is stable. Furthermore, we see that the large $N$ limit is reasonably well satisfied for $\lambda_{max} \leq 0.2$ (i.e., ${M}(\lambda)$ at $\lambda=0$ and at $\lambda \leq 0.2$ are approximately the same).

\begin{figure}[h!]
\centerline{\psfig{figure=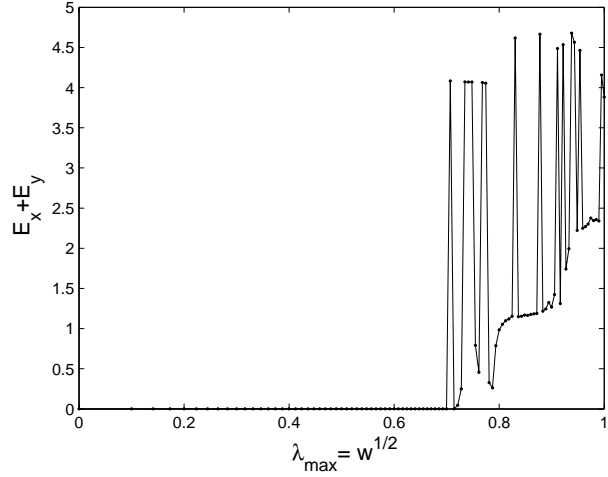,width=9cm}}
\caption{\small The network shown in Fig. \ref{5Nb} is modified as function of $w$. The plot shows the sum of the values of the errors $E_x+E_y$ as function of the corresponding $\lambda_{max}$.   \label{ERRChua}} 
\end{figure}

As a first example, we now consider a specific network where $N$ is small. In particular we consider the network shown in Fig. \ref{5Nb} for which we have shown that the spectrum of $Q$ is given by Eq. (\ref{diciassette}).
Thus $\lambda_{max}=\sqrt{w}$, and the spectrum of $Q$ is real. Fig. \ref{ERRChua} shows the synchronization error at large time as function of $\lambda_{max}$ for $w$ varying between $0$ and $1$ in steps of $0.01$.
We see that, in accord with our stability result from Fig. \ref{MSFChua}, stable synchronization of the chaotic motion is obtained if $\lambda_{max} < 0.7$. In obtaining Fig. \ref{ERRChua}, we initialize the variables $x_i, y_{(1)i}, y_{(2)i}$ randomly in $x_i>0$ on the synchronized chaotic attractor (\ref{synsys}). For these initial conditions, we find that the time asymptotic synchronous motion is on the chaotic attractor of the system (rather than on one of the two fixed point attractors). 

\begin{figure}[h]
\centerline{\psfig{figure=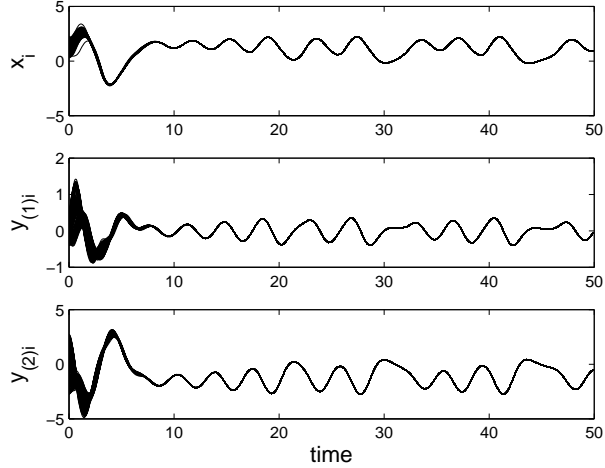,width=9cm}} 
\caption{\small The plot shows the transient evolutions $x_i(t), i=1,..,N_x$ and ${y}_{j(1,2)}(t), j=1,...,N_y$.  The network parameters are as follows: $N_x=200$ and $N_y=300$, $p=0.05$. \label{11b}}
\end{figure}

%


As a second example, Fig. \ref{11b} shows numerical results for a random network (case (i) of Sec. II.C) with $N_x=200, N_y=300, p=0.05$ corresponding to $\lambda_{max}=0.58$, using the same type of initialization as in the previous example.  Since $\lambda_{max} < 0.7$, Fig. \ref{MSFChua} predicts stability, as is in fact seen in Fig. \ref{11b}. In this figure the evolutions of all the randomly initialized systems is plotted versus time. It is seen that good synchronization is achieved by $t \geq 10$.

\section{Networks of Discrete Time systems}

In this section we present a general analysis of two-group, bipartite network synchronization for discrete time systems. We assume the evolution of our discrete time network to be described by the following set of equations:
\begin{equation}\begin{split}
x_i^{n+1}= F(x_i^n) +  \sum_{j=1}^{N_y} A_{ij} H(y_j^n), \quad i=1,...,N_x,  \\
y_j^{n+1}= G(y_j^n) +  \sum_{i=1}^{N_x} B_{ji} L(x_i^n), \quad j=1,...,N_y, \label{Discr1}
\end{split}\end{equation}
where $x_i$ ($y_j$) is a $n_x$ ($n_y$) dimensional vector.
Requiring $A_{ij}$ and $B_{ij}$ to satisfy conditions (\ref{lim1}) and (\ref{lim2}), we see that multi-synchronous motion is possible and is described by the equations,
\begin{equation}\begin{split}
x_s^{n+1}= F(x_s^n) + H(y_s^n),   \\
y_s^{n+1}= G(y_s^n) + L(x_s^n). \label{Discr2}
\end{split}\end{equation}

Linearization of (\ref{Discr1}) about the synchronization manifold leads to:
\begin{equation}\begin{split}
\delta x_i^{n+1}= DF(x_s)\delta x_i^{n}  + DH(y_s) \sum_{j=1}^{N_y} A_{ij} \delta y_j^n, \quad i=1,...,N_x  \\
\delta y_j^{n+1}= DG(y_s)\delta y_j^{n} + DL(x_s) \sum_{i=1}^{N_x} B_{ji}  \delta x_i^n, \quad j=1,...,N_y. \label{Discr2}
\end{split}\end{equation}
 Similar to our previous analysis, we set $\delta x_i^n = c_{x_i} \delta \bar{x}_n$ and $\delta y_j^n = c_{y_j} \delta \bar{y}_n$, where $c_{x_i}$ and $c_{y_j}$ are appropriate scalar coefficients. Substitution of these into (\ref{Discr2}) yields
\begin{subequations} \begin{align}
\delta \bar{x}^{n+1}= DF(x_s) \delta \bar{x}^{n}  + DH(y_s) \sum_{j=1}^{N_y} \frac{A_{ij} c_{yj}}{c_{xi}} \delta \bar{y}^n, \quad i=1,...,N_x  \label{Discr3a} \\
\delta \bar{y}^{n+1}= DG(y_s) \delta \bar{y}^{n} +  DL(x_s) \sum_{i=1}^{N_x} \frac{B_{ji} c_{xi}}{c_{yj}} \delta \bar{x}^n, \quad j=1,...,N_y. \label{Discr3b}
\end{align} \end{subequations}

%
%

Then, following Sec. II, in order for (\ref{Discr3a}) (respectively (\ref{Discr3b}) ) to be satisfied for all $i$ (respectively $j$) we require that 
${(c_{x_i})^{-1}}{\sum_j A_{ij} c_{y_j}}={(c_{y_j})^{-1}}{\sum_i B_{ji} c_{x_i}}= \lambda$, where $\lambda$ is independent of both $i$ and $j$.
 After defining the vector $c=(c_{x_1}, c_{x_2},..., c_{x_{N_x}},c_{y_1}, c_{y_2},..., c_{y_{N_y}})$, the above conditions may be rewritten as
$Q c= \lambda c$ (as in Eq. (\ref{sette})).

This lets us formulate the following master stability function problem
\begin{equation}
\begin{split}
\delta \bar{x}^{n+1}= DF(x_s) \delta \bar{x}^{n}  + \lambda DH(y_s) \delta \bar{y}^n \\
\delta \bar{y}^{n+1}= DG(y_s) \delta \bar{y}^{n} + \lambda DL(x_s) \delta \bar{x}^n. \label{MSFD}
\end{split}
\end{equation}
 As in Sec. II, the master stability function $M(\lambda)$ associates to each (possibly complex) $\lambda_i$ the maximum Lyapunov exponent of the system in (\ref{MSFD}). The synchronous solution is stable if $M(\lambda_i)<0$, for $\lambda_i$ in $\Lambda'$.

\section{More general network topologies}

In this section we consider the case of more general network topologies. Specifically, we remove the constraint that the network is bipartite and we allow connections within the groups. 
We find that stable multi-synchronous evolutions are still possible and can be enhanced when intra-group connections are allowed.

As an Example, we start by considering the following bipartite system,
\begin{equation} \begin{split}
\dot {x}_{i(1)}= -x_{i(2)} - \sum_{j=1}^{N_y} A_{ij} y_{j},& \\
\dot {x}_{i(2)}= 0.2+x_{i(2)} (x_{i(1)}-8.5),& \qquad \qquad i=1,...,N_x; \label{RossN1}
\end{split} \end{equation}
\begin{equation}
\dot {y}_{j}= 0.2 y_{j} + \sum_{i=1}^{N_x} B_{ji} x_{i(1)}, \qquad \qquad j=1,...,N_y, \label{RossN2}
\end{equation}
where $A$ and $B$ satisfy Eqs. (\ref{lim1}) and (\ref{lim2}). In the synchronization manifold Eqs. (\ref{RossN1}) and (\ref{RossN2}) yield the following chaotic R\"ossler system \cite{Rossler},
\begin{equation} \begin{split}
\dot {x}_{s(1)}= -x_{s(2)} -  y_{s},& \\
\dot {x}_{s(2)}= 0.2+x_{s(2)} (x_{s(1)}-8.5),&  \\
\dot {y}_{s}= 0.2 y_{s} +  x_{s(1)}.  \label{RossS}
\end{split} \end{equation}
Assume that the spectrum in $\Lambda'$ includes zero as an eigenvalue. For such a case we see from (\ref{RossN2}) that the $y$-component of the master stability equation (\ref{BLOCK}) yields $\delta \dot{y}= 0.2 \delta y$, giving a Lyapunov exponent of $0.2>0$. Thus the synchronized state is unstable for any network whose spectrum contains zero or small eigenvalues.

\begin{figure}[t]
\centerline{
\psfig{figure=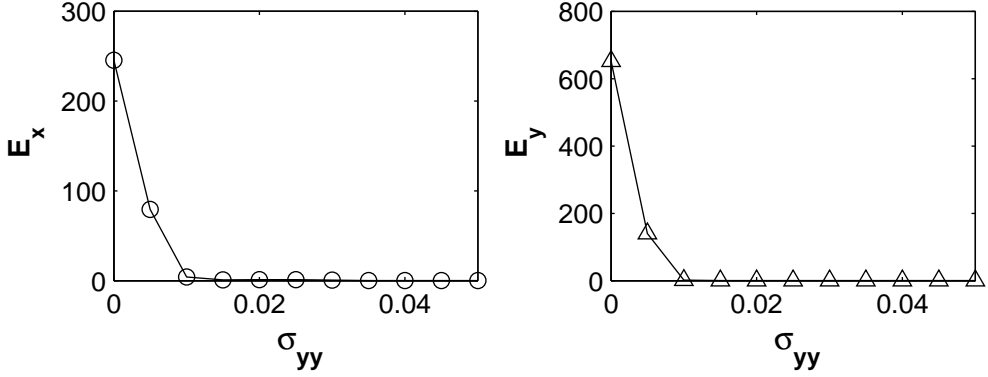,width=15cm} }
\caption{\small The synchronization errors $E_x$ and $E_y$ versus $\sigma_{yy}$ for Eqs. (\ref{Ross1}) and (\ref{Ross2}) for a random network with  $N_x=200$ and $N_y=300$, $p_{xy}=p_{yx}=0.1$ and $p_{yy}=0.15$. \label{DCE}}
\end{figure}

We now ask how this situation is affected by the presence of connections within a group. In order to illustrate this, we consider a case in which the system (\ref{RossN1}),(\ref{RossN2}) is modified by adding connections within the group $S_y$, leading to the following network equations,
\begin{equation}
\begin{split}
\dot {x}_{i(1)}= -x_{i(2)} -  \sum_{j=1}^{N_y} A_{ij} y_{j}& \\
\dot {x}_{i(2)}= 0.2+x_{i(2)} (x_{i(1)}-8.5)& \qquad \qquad  i=1,...,N_x. \label{Ross1}
\end{split} \end{equation}
\begin{equation}
\dot {y}_{j}= 0.2 y_{j} +  \sum_{i=1}^{N_x} B_{ji} x_{i(1)}- \sigma_{yy} \sum_{k=1}^{N_y} \mathcal{L}_{jk} y_k  \qquad \qquad  j=1,...,N_y, \label{Ross2}
\end{equation}
where $\mathcal{L}= \{ \mathcal{L}_{jk} \}$ is a Laplacian matrix: $\sum_k \mathcal{L}_{jk}=0$ for all $j$.
It is important to note that the diffusive coupling term, $\sum_k \mathcal{L}_{jk} y_k$, is null in the synchronization manifold, where the dynamics is governed by the R\"ossler equations (\ref{RossS}).
 In what follows we consider a network with $N_x=200$, $N_y=300$. We generate $A$ and $B$ randomly as described in Sec. II.C, case (i), with $p_{xy}=p_{yx}=0.10$. We generate the Laplacian matrices randomly taking $\mathcal{L}_{jk}$ for $j \neq k$ to be one with probability $p_{yy}$ and $0$ otherwise. Fig. \ref{DCE} shows the effects of varying $\sigma_{yy}$ on the two quantities $E_x$ and $E_y$, defined in Sec. II. We see that the network becomes synchronized for values of $\sigma_{yy}>0.1$, indicating that diffusive intra-group coupling  
 can be effective in enhancing the network synchronization. Note that, as Fig. \ref{DCE} shows, though the diffusive terms are added only to the $y$-systems, synchronization applies for both the systems in $\mathcal{S}_x$ and $\mathcal{S}_y$ (in particular, what is observed is that both the systems synchronize in the chaotic  R\"ossler evolution). However, we wish to emphasize that the master stability function approach presented in Sec. II, is inadequate for assessing the stability of the synchronous evolution when both intra-group and extra-group connections are allowed in the network.

\begin{figure}[t]
\centerline{\psfig{figure=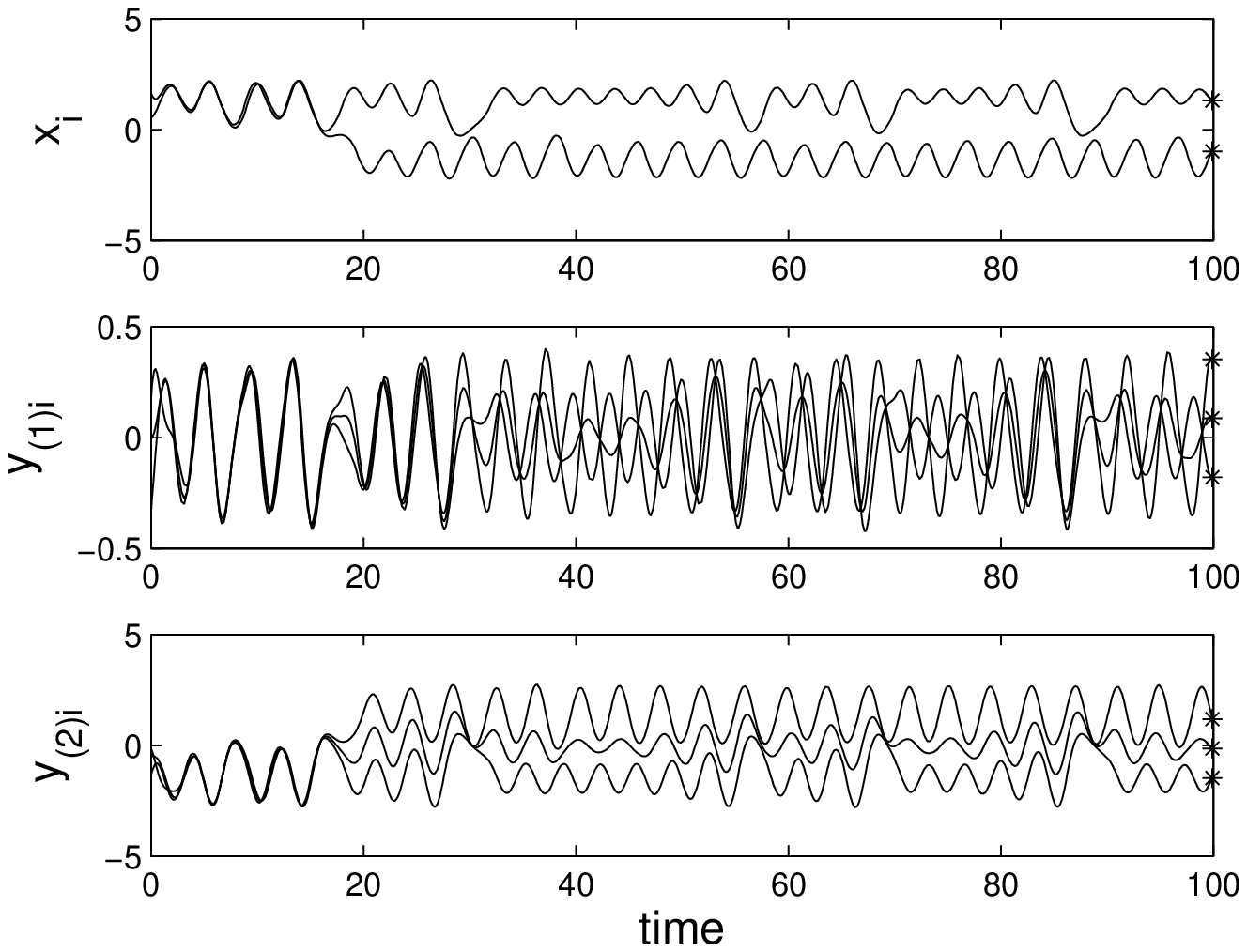,width=9cm} \psfig{figure=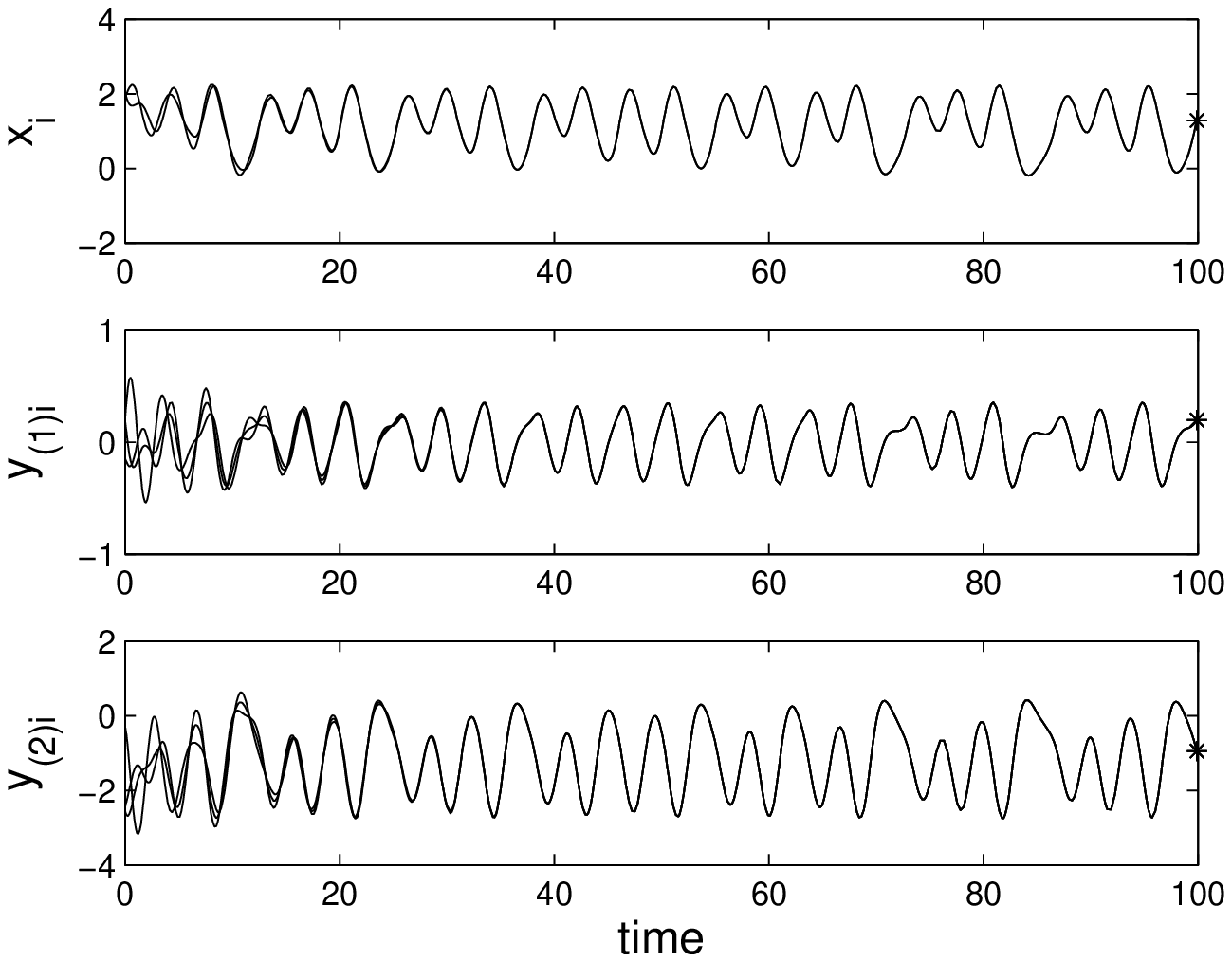,width=9cm}}
\caption{\small The left plots show the time evolutions $x_i(t), i=1,..,2$, and ${y}_{j(1,2)}(t), j=1,...,3$, for the bipartite network represented in Fig. \ref{5Nb} . The equations are those in (\ref{esemcx}), (\ref{esemcy}). The state of the systems at the final time $t^*=100$ is shown by asterisks. The right plots show the state evolution of the network in the case when a bidirectional diffusive link with associated coupling constant equal to $2$ is added between the two $x$-nodes. It is seen that the presence of the added link causes the network to synchronize. \label{FIVE}}
\end{figure}

As another \textit{Example}, we consider the system given by Eqs. (\ref{esemcx}) and (\ref{esemcy}), introduced in Sec. II. The network topology is represented in Fig. \ref{5Nb}, where here we consider the particular case of $w=0.9$.
Thus, since $\lambda_{max}=\sqrt{w}>0.7$, according to the master stability function shown in Fig. \ref{MSFChua}, 
the network is not expected to synchronize. This is indeed what is shown in the left panels of Fig. \ref{FIVE}, where the systems trajectories of the $x$-nodes and $y$-nodes are shown to follow different evolutions.
Now we consider whether it is possible to synchronize the network of these systems by adding diffusive couplings between systems in the same groups.
Namely, we add a single bidirectional diffusive link between the two $x$-nodes and we assume a coupling constant equal to $2$. That is, we add a term $2(x_2-x_1)$ to the right hand side of Eq. (\ref{esemcx}) for $x_1$ and a term $2(x_1-x_2)$ to the right hand side of Eq. (\ref{esemcx}) for $x_2$. As shown in the right panels of Fig. \ref{FIVE}, the network is now observed to synchronize on a multi-synchronous chaotic evolution. In particular, the equations for this evolution are $\dot {x}_s= -r (x_s + h(x_s) )+ r y_{s(1)}, \quad \dot {y}_{s(1)}= -y_{s(1)}+y_{s(2)}+x_s, \quad \dot {y}_{s(2)}= -q y_{s(1)}$ (again, observe that the diffusive coupling term is zero in the synchronization manifold).

\section{Conclusions}

Motivated by the common occurrence in applications of multi-synchronous motions in ensembles of interacting systems characterized by different dynamical behaviors \cite{Ko:Gi98,Jo:Br:Di06,RobotSoccer,Ch:Pa:Ra,Fi:Pa:Sp,Passino,Burt,Ze,Blasius:Stone,Bl:Hu:St,Mo:Ku:Bl,Ryn:Loy,St:Ol:Bl:Hu:Ca,Maynard}, we have addressed the issue of how these systems can stabilize in distinct (possibly chaotic) synchronous evolutions. 
This form of synchronization is distinct from both diffusive coupling synchronization \cite{Pe:Ca} and replacement synchronization \cite{Replace}.

By considering the 
underlying network of connections among the systems, we report conditions for the existence of a synchronization manifold.  In the case of bipartite network topologies (i.e., when there are two communities and network links only connect nodes in different communities) we studied the stability of the synchronization manifold by means of a master stability function approach. In so doing, it was possible to decouple the effects of the network topology from those of the dynamics at the network nodes. We also presented an extension of our approach to discrete time systems.

Finally, we considered examples of the case of more general network topologies, where links are also allowed to fall within each community, and we reported numerical evidence that the presence of diffusive couplings among nodes within the same community can enhance the network synchronizability \footnote{We were able to carry out a master stability analysis only in the particular case where the network had two groups and was assumed to be bipartite.}. 

We believe this paper represents only a first step in the study of multiple synchronization of complex networks. We hope that our work will stimulate further research efforts to address this issue in the future.

This work was supported by ONR (Physics), NSF (PHY0456249), and by a MURI contract administered by ONR.


\end{document}